\documentclass[reprint,aps,prl,nofootinbib,twocolumn,superscriptaddress,preprintnumbers]{revtex4-1}

\usepackage{tensor}

\usepackage{dcolumn}
\usepackage[
	  pagebackref=false,
	  colorlinks=true,
      linkcolor=blue,
      urlcolor=blue,
      filecolor=black,
      citecolor=red,
      pdfstartview=FitV,
      pdftitle={},
        pdfauthor={},
        pdfsubject={},
        pdfkeywords={},
        pdfpagemode=None,
        bookmarksopen=true
      ]{hyperref}

\usepackage[normalem]{ulem}
\usepackage{lipsum}
\usepackage{amsthm}
\usepackage{amsmath}
\usepackage{enumerate}
\usepackage{amsfonts}
\usepackage{epsfig}
\usepackage{mathbbol}
\usepackage{mathrsfs}

\usepackage{xcolor}

\usepackage{amssymb,graphicx,mathtools}

\newcommand{\dif}{\mathrm{d}}
\newcommand{\pa}[1]{\left(#1\right)}

\newcommand{\xtd}[1]{e^{#1}_{(X)}}
\newcommand{\ytd}[1]{e^{#1}_{(Y)} }

\newcommand{\atan}{\text{atan}}

\def\d{\delta}
\def\D{\Delta}
\def\f{\frac}

\def\m{\mu} 
\def\n{\nu}

\def\p{\phi}

\def\t{\theta}

\def\be{\begin{equation}}
\def\ee{\end{equation}}
\def\bag{\begin{aligned}}
\def\eag{\end{aligned}}
\def\bea{\begin{eqnarray}}
\def\eea{\end{eqnarray}}
\def\ba{\begin{array}}
\def\ea{\end{array}}

\def\bc{\begin{center}}
\def\ec{\end{center}}

\begin{document}

\title{Black Hole Ringdown Seen in Photon Polarization Swings}

\author{Jiewei Huang$^{\ddag}$}
\affiliation{Department of Physics, Peking University, No.5 Yiheyuan Rd, Beijing
100871, P.R. China}

\author{Yehui Hou$^{\ddag}$}
\affiliation{Tsung-Dao Lee Institute, Shanghai Jiao-Tong University, Shanghai, 201210, P. R. China}

\author{Zhen Zhong}
\affiliation{Dipartimento di Fisica, Sapienza Università di Roma, Piazzale Aldo Moro 5, 00185, Roma, Italy}
\affiliation{INFN, Sezione di Roma, Piazzale Aldo Moro 2, 00185, Roma, Italy}

\author{Minyong Guo}
\email{Corresponding author: minyongguo@bnu.edu.cn}
\affiliation{School of physics and astronomy, Beijing Normal University, Beijing 100875, P. R. China}
\affiliation{Key Laboratory of Multiscale Spin Physics (Beijing Normal University), Ministry of Education, Beijing 100875, China}

\author{Bin Chen}
\email{Corresponding author: chenbin1@nbu.edu.cn}
\affiliation{Department of Physics, School of Physical Science and Technology, Ningbo University, Ningbo, Zhejiang 315211, China}
\affiliation{Department of Physics, Peking University, No.5 Yiheyuan Rd, Beijing
100871, P.R. China}
\affiliation{Center for High Energy Physics, Peking University, No.5 Yiheyuan Rd, Beijing 100871, P. R. China}

\begingroup
\renewcommand\thefootnote{\ddag}
\footnotetext{Co-first authors}
\endgroup

\begin{abstract}

Light propagating through a perturbed spacetime could imprint the underlying gravitational waveform directly onto electromagnetic observables. In this Letter, we develop a covariant perturbative framework for polarized photon propagation in generic curved spacetimes, and derive a compact expression for the observable polarization-angle (PA) swing during Kerr ringdown, explicitly demonstrating its time-domain locking to the quasi-normal modes. We confirm this behavior using dynamical ray-tracing calculations for a broad class of photon trajectories.
Photons grazing the strong-field region exhibit an achromatic, damped PA oscillation that tracks the ringdown, with a phase set by the mode’s angular structure. The swing amplitude can reach $\sim 10^{\circ}$ and leaves distinctive signatures in spatially resolved autocorrelations.
These results open a new polarimetric window onto black hole mergers and ringdown.

\end{abstract}
\maketitle

\noindent \emph{\textbf{Introduction}}---
Compact-object mergers and black-hole formation provide a unique arena for testing gravity in the dynamical strong-field regime. Of particular interest is the post-merger ringdown, whose quasinormal-mode (QNM) spectrum encodes the properties of the remnant black hole \cite{Berti:2009kk, London:2014cma, Cardoso:2016rao, Isi:2019aib, Giesler:2019uxc, Bhagwat:2019bwv}, and is now being probed with increasing precision by current and future gravitational-wave (GW) detectors \cite{LIGOScientific:2016aoc, LIGOScientific:2020ibl, KAGRA:2021vkt, LIGOScientific:2025rid, Punturo:2010zz, TianQin:2015yph, Hu:2017mde, LISA:2017pwj, Reitze:2019iox}. This progress naturally raises a broader question: can black-hole ringdown also leave observable imprints outside the GW channel itself\,?

One possible route is through photon polarization swings. Gravitational perturbations affect not only the trajectories of light rays but also the transport of their polarization vectors \cite{skrotskii1957influence, plebanski1960electromagnetic, schneider2005gravitational}.
Beyond the geometric reorientation of polarization plane associated with light bending, spacetime curvature also causes the polarization-vector rotation within the plane (gravitational Faraday rotation, GFR) \cite{balazs1958effect, Mashhoon:1975ki, Faraoni:1992jb, Prasanna:2001xk, Faraoni:2007uh}.
Near a ringing black hole, strong lensing bends light rays to efficiently sample the near-horizon region where QNM perturbations are strongest \cite{Zhong:2024ysg}, enabling horizon-scale metric dynamics to imprint a significant GFR on the PA. At larger radii, photon trajectories tend to become collinear with the outgoing perturbations, suppressing dephasing and limiting phase averaging.
Consequently, the near-horizon GFR can survive to infinity, potentially producing a time-dependent oscillatory PA signal tied to the ringdown.

This possibility becomes especially relevant in astrophysical scenarios where polarized light adequately probes the strong-field region. Such emission may arise from post-merger accretion flows \cite{Palenzuela:2009yr, Bode:2009mt, Metzger:2011bv, Farris:2012ux, Giacomazzo:2012iv, Gold:2014dta, Perna:2016jqh, Gutierrez:2021png, Ennoggi:2025ijc}
or some localized hotspots \cite{Broderick:2005my, Ripperda:2021zpn, Wielgus:2022heh}, while highly polarized background emitters, such as compact synchrotron-emitting regions in AGN jets (e.g., bright jet knots) \cite{saikia1988polarization, marscher2008inner} or, in favorable alignments, background blazars \cite{angel1980optical, impey1990optical}, can also be strongly lensed through the near-horizon region.
Although the surrounding plasma induces chromatic Faraday rotation, the gravitationally induced GFR is achromatic and spectrally separable. Sourced by QNM ringdown, it would additionally show a characteristic damped oscillatory time dependence.
PA swings thus offer a potentially complementary electromagnetic probe of black-hole perturbations, alongside GW measurements and astrometric observables \cite{Zhong:2024ysg}.

Motivated by this prospect, in this Letter we study photon-polarization evolution in the strong-field region of a ringing black hole and the resulting polarimetric signatures. We derive a general expression for the observable PA variation in perturbation theory and evaluate it numerically in a dynamical black-hole background constructed within linear perturbation theory. 
For photons emitted both near the hole and by distant background sources, we find that the time-resolved PA exhibits a damped oscillation tied to the QNM frequency, while phase differences across the image encode the angular structure of the underlying mode. These results establish polarization as a promising tracer of black-hole ringdown.
We use units $G=c=1$ throughout.


\medskip

\noindent {\it \textbf{Covariant Perturbative Description}}--- 
We first develop a covariant description of polarized light propagation in a generic dynamical spacetime, accurate to first order in metric perturbations. The framework is broadly applicable to GW environments, including compact-binary inspiral, stochastic backgrounds, and black-hole relaxation \cite{regge1957stability, vishveshwara1970stability, kokkotas1999quasi, Maggiore:1999vm, Berti:2009kk, Christensen:2018iqi, Barack:2018yvs}.
To isolate the imprint of gravitational perturbations, we assume that photons propagate through vacuum after emission.

In the geometric-optics limit, where the photon wavelength is much shorter than the local curvature scale, the polarization state is described by the Stokes vector $\mathcal{S} = \{\mathcal{I}, \mathcal{Q}, \mathcal{U}, \mathcal{V}\}$. These quantities are defined in the radiation gauge with respect to an orthonormal tetrad $\{ e^\mu_{(X)}, e^\mu_{(Y)} \}$ spanning the screen space orthogonal to the wave vector \cite{Broderick:2003bg, Broderick:2003fc, Shcherbakov:2010kh}. Along each null ray, the Stokes parameters are conserved, \(\mathrm{d}\mathcal{S}/\mathrm{d}\lambda = 0\), with \(\lambda\)  an affine parameter, while the polarization basis is parallel transported according to \(k^\nu\nabla_\nu e^\mu_{(i)}=0\) with $k^\mu$ the tangent vector. The GFR is thus encoded entirely in the transport of this transverse tetrad.
  
In a perturbed spacetime, the metric is split into the background term \(g_{\mu\nu}\) and a small perturbation \(h_{\mu\nu}\). 
Photons emitted with the same initial polarization can then experience different GFRs due to the perturbation. 
To quantify this difference, we introduce the perturbed polarization rotation (PPR) \(\vartheta\), defined as the change in polarization angle relative to the unperturbed background. 
Explicitly,
\bea\label{eq:deftheta}
\vartheta = \Theta_{\mu\nu}\, e^\mu_{(X)}e^\nu_{(Y)},
\eea
where the tetrad is defined in the background spacetime, and \(\Theta_{\mu\nu}\) is an antisymmetric tensor that characterizes the local Lorentz rotation between the tetrads in the perturbed and background spacetimes. 
Eq.~\eqref{eq:deftheta} thus defines a covariant generalization of that on a flat background \cite{Prasanna:2001xk, Kopeikin:2001dz, Kopeikin:2006gf, Shoom:2022oer}.
Physically, \(\vartheta\) represents the cumulative rotation of the polarization plane as measured by an observer who parallel-transports a reference frame along the background geodesic.
The evolution of \(\Theta_{\mu\nu}\) along the ray is given by
\bea\label{eq:polarderv2form}
&&k^\alpha\nabla_\alpha \Theta_{\mu\nu} = R_{\mu\nu\alpha\beta} k^\alpha \xi^\beta + k^\rho \nabla_{[\mu} h_{\nu]\rho}, \\
&&k^\alpha\nabla_\alpha(k^\beta\nabla_\beta \xi^\mu) = R^\mu_{\ \nu\alpha\beta} k^\nu k^\alpha  \xi^\beta 
 - \tensor{H}{^\mu_\nu_\rho}k^\nu k^\rho\,,\label{eq:geoder}
\eea
where the derivatives are taken along the unperturbed geodesic, and \(R_{\mu\nu\alpha\beta}\) is the background Riemann tensor;
\(\xi^\mu\) is the geodesic deviation vector describing the displacement from the background trajectory, and $ \tensor{H}{^\mu_\nu_\rho}\equiv g^{\mu\sigma}
    \pa{
        \nabla_{\nu}h_{\sigma\rho}+\nabla_{\rho}h_{\nu\sigma}-\nabla_{\sigma}h_{\nu\rho}
    }/2$. 
The first terms on the right-hand sides of Eqs.~\eqref{eq:polarderv2form} and \eqref{eq:geoder} describe the effect of background curvature acting through the geodesic deviation, while the second terms are source terms from the metric perturbation. 
A detailed derivation of \(\vartheta\) and \(\Theta_{\mu\nu}\) is provided in Supplemental Material \ref{App:A} and \ref{App:B}.

Substituting the definition of \(\vartheta\) into the evolution equation for \(\Theta_{\mu\nu}\) yields a complete, gauge-invariant description of polarization rotation. 
Notably, the rate of change of \(\vartheta\) is independent of the initial polarization state, generalizing known results in flat spacetime \cite{Liang:2025vji}. 
In the regime where the perturbation-induced term dominates over the background curvature contribution, 
the total PPR accumulated along the ray from the initial emission position \(x_s^\mu\) to the observer \(x_o^\mu\) simplifies to
\bea\label{eq:thetaevo}
\vartheta(x_s^\mu,x_o^\mu) \simeq \int_{x^\mu(\lambda)} \Omega\left(x^\mu\right)\mathrm{d}\lambda\,,
\eea
where we have introduced the local rotation rate
\bea\label{eq:Omegadef}
\Omega\left(x^\mu\right) \equiv e^\mu_{(Y)}e^\nu_{(X)}\, k^\rho \nabla_{[\mu} h_{\nu]\rho},
\eea
which covariantly describes the instantaneous polarization rotation at the spacetime point \(x^\mu(\lambda)\). It is a curved-spacetime generalization of the angular velocity introduced in \cite{Kopeikin:2001dz}.
Eq.~\eqref{eq:thetaevo} provides a direct prescription for computing the evolution of the PA using the background geodesic equation \(k^\nu\nabla_\nu k^\mu=0\), together with a local rotation rate defined by a linear operator acting on $h_{\mu\nu}$. Integrating the PPR along the trajectory to the observer then yields the correct time-dependent PA evolution.

\medskip

\noindent {\it \textbf{Imprint of Kerr Ringdown}}---
As a prototypical astrophysical process, we apply the polarization-analysis framework to a perturbed Kerr spacetime to explicitly show the ringdown imprint.
The radiative degrees of freedom of the Kerr QNM are encoded in the perturbations of the Weyl scalars $\Psi_0$ and $\Psi_4$, typically defined with respect to the Kinnersley tetrad in the Boyer-Lindquist coordinate $\{t,r,\t,\phi\}$ \cite{kinnersley1969type}. The corresponding master variables, $\psi_2 = \Psi_0$, $\psi_{-2} = \left(r - i a \cos{\t}\right)^4\Psi_4$, satisfy the separable Teukolsky equation with spin weight $s = \pm 2$ and admit solutions of the form \cite{Teukolsky:1973ha}
\bea\label{eq:RDphase}
\psi_s = \f{_{s}Z}{\sqrt{2\pi}}\, R_s(r) \,S^{lm}_s(\t) \, e^{-i \omega t + i m \phi} \,.
\eea
The metric perturbation $h_{\mu\nu}$ is then reconstructed from $\psi_{\pm 2}$ using the Chrzanowski-Cohen-Kegeles (CCK) formalism, in which $h_{\mu\nu}$ is obtained by applying a second‑order differential operator to the associated Hertz potentials \cite{Chrzanowski:1975wv,Cohen:1974cm,Kegeles:1979an}. 
To enforce causal GW propagation, we impose a Heaviside-like filter along the outgoing null direction via $h_{\mu\nu} \to H(t - r_*)h_{\mu\nu}$, where $r_*$ is the tortoise coordinate \cite{Bardeen:1972fi}. 
This ensures that the perturbation is confined within the light cone $t = r_*$ \cite{Leaver:1986gd, kokkotas1999quasi}. 

\begin{figure}[!htp]
    \centering
    \includegraphics[width=1\columnwidth]{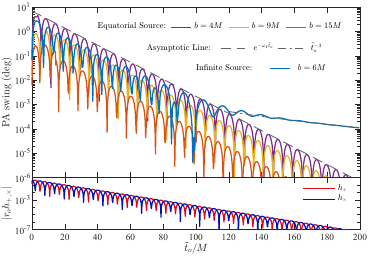}
 \hspace*{-3mm} \includegraphics[width=1.01\columnwidth]{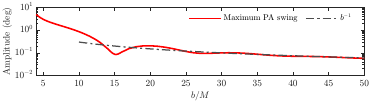}
     \caption{Behavior of the photon polarization and GW polarization as received by a distant on-axis observer, shown as functions of the observation time $\tilde{t}_o$. We consider a Kerr black hole with spin $a = 0.7 M$. The ringdown is dominated by the fundamental tone with $\{n,l,m\}=\{0,2,2\}$ and complex frequency $M\omega = 0.5326 - 0.08079\,i$ \cite{Berti:2009kk}. The magnitude of the outgoing mode is set to $|_{-2}Z| = 0.1$. 
     \textbf{Top}: PPR of photons emitted from the equatorial plane or from spatial infinity in the Kerr spacetime. Colors denote rays with the same polar angle $\varphi=0$ but different radius $b$ on the observer's screen. Exponential and power-law reference curves (dash-dotted) are overplotted to characterize the temporal evolution of the PA.
\textbf{Middle}: The two GW polarization components, $h_+$ and $h_\times$, obtained via transverse projection onto the observer's screen. 
\textbf{Bottom}: Peak PPR amplitude versus polar radius $b$.}
     \label{fig:pa}
\end{figure}

During the ringdown phase, the PPR is dominated by the GW-induced source term in Eq.~\eqref{eq:polarderv2form}, so Eq.\eqref{eq:thetaevo} provides a good estimate. This reflects the different radial scalings: the background curvature falls off as $ r^{-3}$ and is localized near the horizon, while the geodesic deviation $\xi^{\mu}$ grows mainly at large $r$; by contrast, the ringdown perturbation scales as $h \sim r^{-1}$ and thus dominates the PPR evolution \footnote{We have checked that, for most light rays, the GW-induced source term exceeds the curvature term by at least an order of magnitude, consistent with the good agreement between Eq.\eqref{eq:RAexpression} and the numerical results shown in Fig.\ref{fig:pa}.}.
Meanwhile, owing to the stationarity and axisymmetry of the Kerr background, the local rotation rate in Eq.~\eqref{eq:Omegadef} admits a separable form during ringdown, $\Omega\pa{x^\mu}=\Re[\tilde{\Omega}\pa{r,\t} e^{i\left(m\p-\omega t\right)}]\cdot H\pa{t-r_*}$.
Based on these considerations, we derive a compact expression for the PA swing measured by a distant static observer. The resulting signal exhibits a damped harmonic form that exactly samples the QNM waveform:
\begin{align}\label{eq:RAexpression}
    \vartheta\pa{\tilde{t}_o}=|\mathcal{A}| \cos\pa{\omega_R\, \tilde{t}_o - \Phi}e^{-\omega_I\tilde{t}_o}\,,
\end{align}
where $\tilde{t}_o$ denotes the photon arrival time at the observer \footnote{Hereafter we redefine the time as $\tilde{t}_o = t_o - (r_*)_o$ such that $\tilde{t}_o = 0$ corresponds to the initial arrival time of the gravitational wave.}, and $\omega_R$, $\omega_I$ are the real and imaginary parts of the QNM frequency, respectively. 
Eq.~\eqref{eq:RAexpression} applies to both the fundamental mode and its overtones, and captures one of the central results of this Letter: the evolution of the observed polarization angle is effectively ``frozen into'' the ringdown waveform, with its synchronization set by the ringdown dynamics. Strictly speaking, this expression is derived under the idealized assumption that the photon experiences the perturbation along its entire post-emission trajectory. Nevertheless, as shown below, it accurately describes a much broader class of PPR behavior, particularly for photons perturbed mainly during their outward propagation from the horizon.
The amplitude coefficient $\mathcal{A} \equiv \int_{\lambda_s}^{\lambda_o}   \dif\lambda\,\tilde{\Omega} \pa{\lambda}e^{i\left(m\p(\lambda) -\omega t(\lambda)\right)}$ depends on the endpoint positions, and varies across photon trajectories. The full derivation is presented in Supplementary Material \ref{app:timePPR}. 

\medskip
\noindent {\it \textbf{Dynamical Polarimetric Signatures}}---
To provide a concrete visualization of how the time-evolving PA encode the ringdown, we develop a numerical backward ray-tracing scheme in the dynamical spacetime that efficiently solves the full set of null geodesics and parallel transport equations for polarization, yielding the PPRs across the image plane for arbitrary spins, emission and viewing geometries. The numerical approach is in agreement with the analytical perturbative prediction (see Supplementary Material \ref{app:nveps}).

Here we consider two representative emission geometries: (i) rays connecting the black hole's equatorial plane to the observer, and (ii) rays originating from spatial infinity that reach the observer after grazing the black hole. The former corresponds to the accretion disk emission, while the latter models background emission from distant sources.
Strong lensing imprints these two geometries differently \cite{Virbhadra:2008ws, Cunha:2018acu, Gralla:2019drh, Gelles:2021kti}, and we expect their signals to exhibit different responses to GWs. 
The observer's image plane is spanned by the polar coordinates $(b, \varphi)$, tailored to imaging in dynamical spacetimes \cite{Zhong:2024ysg}. 
For clarity, we restrict the main text to an observer at the south pole of the black hole. Results for inclined viewing configurations are qualitatively similar and are deferred to the Supplementary Material \ref{sec:dfhhdj}.

Fig.~\ref{fig:pa} shows the time evolution of the PA on the image plane versus the observation time $\tilde{t}_o$, for different photon incidence directions parameterized by $(b, \varphi)$, for the fundamental mode of a Kerr black hole with spin $a = 0.7 M$. 
Here, we subtract the time-independent background PA value.
For the equatorial-plane emission, the PPR exhibits an long-term damped oscillation $\propto e^{-\omega_I \tilde{t}_o}$ as predicted by Eq.~\eqref{eq:RAexpression}, closely tracking the ringdown waveform over a long timescale $\tilde{t}_o \approx 200 M$.
This behavior arises because the accumulated imprint along each trajectory is determined primarily by the initial GW phase at equatorial emission, effectively confining the trajectories to constant-phase surfaces and causing the PA evolution to track the QNM waveform.
This implies that, once emitted near the horizon, the photon polarization swing can be frozen into the ringdown pattern, thereby encoding the essential features of the GW signal.

For photons propagating from spatial infinity toward the black hole, the PA shows distinct time dependence for early- and late-arriving trajectories. Early-arriving photons are perturbed mainly during the outgoing phase, leading to the freezing behavior predicted by Eq.\eqref{eq:RAexpression}, as in the case of photons emitted from the equatorial plane. By contrast, late-arriving photons ($\tilde{t}_o \gtrsim 100M$) are affected primarily during the ingoing phase, before entering the near-horizon region. Because the ringdown damps rapidly, the dominant contribution to the PPR accumulates near $t \simeq r_*$. In this regime, the photon wave vector is nearly aligned with the GW propagation direction, so the polarization evolution is governed mainly by nonspherical components. After averaging over the rapid GW oscillations, the PPR depends only on the mean GW intensity, which decays as $r^{-3}$ \cite{Kopeikin:2001dz}. Consequently, $\vartheta \simeq (\tilde{t}_0)^{-3}$, as shown in Fig.~\ref{fig:pa}, consistent with the deflection-angle variation \cite{Damour:1998jm, Zhong:2024ysg}. 

The envelope amplitude of $\vartheta$ reflects the strength of the gravitational perturbation near the emission point.
The QNM profile peaks near the prograde unstable photon orbit \cite{ferrari1984new, schutz1985black}
located at $r_{\rm ph} \approx 2.07 M$ for $a = 0.7 M$. This corresponds to a critical impact parameter $b_c \approx 3M$ for photons reaching the observer directly \cite{Gelles:2021kti}, \footnote{For photons emitted from the equatorial plane and escaping directly to an on-axis observer without undergoing orbital winding, the source coordinates $(r_s,\phi_s)$ are simply related to the screen coordinates $(b,\varphi)$ through \cite{Gates:2020sdh, Chen:2024jkm}:
$r_s \approx b - 1 + {(1 - a^2)}/{2b}$, $\phi_s \approx \varphi + {2a}/{r_s^2}$. Although derived in  large-$b$ regime, they provide good approximations to the mappings at radii of a few $M$.}  where the PA swing attains its maximum amplitude.
As $b$ moves away from $b_c$, the swing amplitude decreases, as shown by the red, yellow, and blue curves in Fig.~\ref{fig:pa}. Notably, the peak PA swing reaches $\vartheta_{\rm max}\sim 10^{\circ}$ near $\tilde{t}_o = 0$ for $b = 4M$, indicating a substantial polarimetric imprint.
Note that we fix the QNM radiation energy to be $E_{\rm tot}\simeq 0.6\%\,M$ as a conservative choice. For larger values, e.g., the estimated maximum $E_{\rm tot}\simeq 3 \%\,M$ \cite{Berti:2007fi}, $\vartheta_{\rm max}$ scales correspondingly. 
For on-axis viewing, we also find an approximate far-zone scaling, $\vartheta_{\rm max} \simeq 40^{\circ}\sqrt{E_{\rm tot}/M}\left(b/M\right)^{-1}$ for $b \gg M$. This trend arises because light is most affected when its wave vector is approximately perpendicular to the GW direction, where the spherical component dominates and follows a $r^{-1}$ falloff. Near the black hole, the nontrivial radial profile of the QNM produces a more intricate peak-swing structure, as illustrated in the bottom panel of Fig.~\ref{fig:pa}.

A detectability criterion for PA swings in a given merger can be formulated as follows. For high signal-to-noise observations,  the polarization-angle uncertainty is $\sigma_\text{PA} \sim (P\lambda \sqrt{I t_{\text{e}}})^{-1}$, where $I,P$ are the surface brightness and polarization degree, $\lambda$ is the observing frequency, and $t_{\text{e}}$ is the exposure time \cite{howell2000handbook}. Resolving the ringdown-induced PA swings then requires
\bea\label{eq:detect}
\lambda P \sqrt{I t_e} \,\gtrsim \, 0.048 \f{b}{M} \f{\left(1+q\right)^2}{q} \,,
\eea
where we adopt the total emitted GW energy $E_{\rm tot} \sim 0.27 Mq^2(1+q)^{-4}$ with $q$ the binary mass ratio \cite{Berti:2007fi}. Note that this estimate is conservative, as it is based on the far-zone expression, which underestimates the amplitude relative to that near the horizon.

\begin{figure}[!htp]
    \centering
    \includegraphics[width=1\columnwidth]{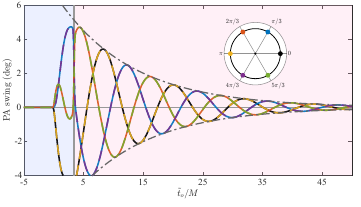}
    \caption{PA swing as a function of the observer time $\tilde{t}_o$, shown for fixed $b = 4 M$ and different polar angles: $\varphi = 0$, $\pi/3$, $2\pi/3$, $\pi$, $4\pi/3$, $5\pi/3$. The emission source is located in the equatorial plane. 
The spacetime setup is the same as in Fig.~\ref{fig:pa}. The dash-dotted curves indicate the envelope of $\vartheta$. Because the GW perturbation requires a finite time to reach the emission point, the PPR is not fully governed by the ringdown at very early times (blue-shaded region). It is then rapidly captured by the QNM and becomes fully synchronized for $t \gtrsim 5M$ (shaded red region).}
    \label{fig:mdcx}
\end{figure}

Because the phase of a single-mode ringdown scales with the azimuthal angle as $e^{im\phi}$, the PA phase difference between rays reaching the image plane at different polar angles is shaped by the angular structure of the QNM. For on-axis viewing, we find that the phase $\Phi$ in the frozen-in pattern Eq.~\eqref{eq:RAexpression} reduces to a simple form, which directly encodes the QNM azimuthal number:
\bea\label{eq:faceonPhi}
\Phi = m\,\varphi + \Phi_0 \,,
\eea
where the constant $\Phi_0$ can be removed through coordinate redefinition. A rotation $\varphi\to \varphi + \d$ of the image plane correspondingly rotates the light path, leading to a QNM phase shift of $m\d$ along the trajectory, which is then directly imprinted on $\vartheta$. 
To better illustrate this, in Fig.~\ref{fig:mdcx} we present the temporal evolution of $\vartheta$ driven by the fundamental mode. Although the emission is initially launched in the equatorial plane, it exhibits only small deviations from the infinite-source limit for $\tilde{t}_o \leq 100 M$.
The received photons have the same radius $b = 4M$, but different polar angles. The PA difference between them demonstrates the breaking of axisymmetry during the ringdown.  
For pairs of angles separated by $\pi$, the evolution of $\vartheta$ follows similar patterns, reflecting the characteristic angular structure of the $m=2$ mode. 
Therefore, if the source can be spatially resolved on the image plane, the PPR may provide a direct probe of the angular distribution of the gravitational perturbation.

\medskip
\noindent {\it \textbf{Autocorrelation analysis}}---
Gravitationally induced photon signatures can be further characterized using statistical image observables that isolate quasi-periodic signals from astrophysical noise \cite{Hadar:2020fda, Chen:2022kzv, Hadar:2023kau, Zhang:2025vyx, Bezdekova:2025wii}. 
To this end, we introduce the image-resolved autocorrelation function (ACF) of the PA ($\vartheta_o$) between two image pixels $\vec{x},\vec{y}$,
\bea\label{eq:defco}
C(\vec{x},\vec{y},t) = \f{1}{c_0} \int_{-\infty}^{\infty} \vartheta_o(\vec{x},t') \,\vartheta_o(\vec{y}, t' + t)\,dt' \,,
\eea
where $c_0$ is a normalization constant. As a representative case, taking $\vec{x}$ on the horizontal axis, one obtains, for the frozen-in mode in Eq.~\eqref{eq:RAexpression} viewed on-axis, a damped harmonic form
\bea\label{eq:acf01}
C = \f{|\mathcal{A}_{x}||\mathcal{A}_{y}|}{4\,\omega_I c_0}\cos{\left(  \omega_R t + m \varphi_y -\phi' \right)}\, e^{-\omega_I t}  \,,
\eea
where $\phi' = \arctan{\left[2\omega_R\omega_I ( \omega_R^2+2\omega_I^2 )^{-1}\right]}$. This form makes the physics transparent: the time lag tracks the QNM frequency and damping, while the angular lag isolates the azimuthal number $m$. 

\begin{figure}[!htp]
\centering
    \includegraphics[width=1\columnwidth]{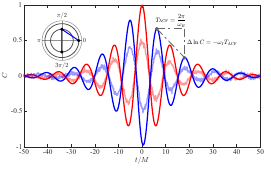}
     \caption{Two-point ACFs of PAs in the time-lag domain, for two representative configurations. 
Solid curves correspond to the noiseless simulated signal, while translucent curves show the results after adding Gaussian white noise. The noise standard deviation is chosen to give a signal-to-noise ratio of zero. The spacetime setup is the same as in Fig.~\ref{fig:pa}.}
     \label{fig:correlation}
\end{figure}

In Fig.~\ref{fig:correlation}, we compute the ACFs for two representative pairs of image locations driven by the fundamental mode: $(b, \varphi) = (4M,\pi/2)$ and $(4M,3\pi/2)$ (red curve), and correlation between 
$(b, \varphi) = (4M,3\pi/2)$ and $(6M,0)$ (blue curve).
In both cases, the ACF exhibits quasi-periodic oscillations with a period set by the real part of the QNM frequency, $T_{\rm ACF} = 2\pi/\omega_R \approx 11.8 M$. The temporal correlation length is controlled by the damping rate, yielding a logarithmic decrement per cycle of $\D \ln C = -\omega_I T_{\rm ACF} \approx  -0.95$. The two ACFs are phase-shifted by $\approx \pi$, consistent with the $m =2$ structure.

The white-noise injection test (translucent curves in Fig.~\ref{fig:correlation}) shows that the PA signal is robust against short-correlation perturbations. ACF-based analyses thus suppress broadband stochastic variability, including plasma-kinetic fluctuations \cite{Ball:2018icx, Comisso:2019frj, Ripperda:2021zpn}. Yet the accretion flow itself varies on similar timescales: orbital motion near the ISCO and MHD turbulence evolve on dynamical timescales of order $M$ \cite{Hawley:1995sy, balbus1998instability, uttley2005non, GRAVITY:2020lpa}, potentially contaminating the ringdown signal. Isolating the latter will likely require joint modeling of the damped transient and the correlated astrophysical background \cite{finn1992detection, cutler1994gravitational, Berti:2005ys}. Multiband observations may further help distinguish the two, as gravitationally driven responses are nearly achromatic, unlike the chromatic variability of the accretion flow.

\medskip

\noindent \emph{\textbf{Summary and Discussion}}--- 
We have developed a covariant, source-independent framework for describing how gravitational perturbations modulate null rays and their linear polarization. Within this framework, we show that the induced PA swing can become imprinted onto the QNM waveform, thereby encoding the ringdown of a Kerr black hole (Eq.\eqref{eq:RAexpression}). Our numerical calculations, performed over a broad range of photon trajectories, support this behavior and underscore its potential relevance for black-hole polarimetry. Specifically, the period of the PA oscillation tracks the real part of the QNM frequency, its damping rate follows the imaginary part (Fig.\ref{fig:pa}), and phase offsets across the image plane encode the azimuthal structure of the mode (Fig.\ref{fig:mdcx}). With peak amplitudes of order $10^{\circ}$, this effect may be sufficiently large to motivate observational searches in favorable systems. Combined with correlation-based strategies for disentangling it from intrinsic astrophysical variability (Fig.\ref{fig:correlation}), the PA swing emerges as a promising geometric probe of black-hole ringdown.
 
Our analytical arguments suggest that the sensitivity of PA evolution to gravitational perturbations is not restricted to the late-time ringdown, but may extend across a broader range of dynamical regimes. In particular, during the early ringdown, where higher overtones contribute \cite{Leaver:1985ax, Giesler:2019uxc, Isi:2019aib}, the superposition of multiple quasinormal modes may imprint a characteristic, mode-dependent structure on the observed polarization signal.  
Moreover, because our framework does not rely on any specific emission geometry, it applies generically to a wide class of polarized sources, ranging from millimeter synchrotron emission to X-ray systems \cite{connors1980polarization, Schnittman:2009im, Schnittman:2009pm, bellazzini2010x, Krawczynski:2012ac, EventHorizonTelescope:2021bee, GRAVITY:2020lpa, Wielgus:2022heh, EventHorizonTelescope:2024hpu}.

For equatorial emission, the frozen-in pattern persists to at least $200\,GM/c^3$ (Fig.~\ref{fig:pa}). Although its amplitude decays exponentially, this highlights the role of post-merger accretion flows in producing observable polarization-angle (PA) swings. Such signals can arise from accreting, hot, magnetized plasma \cite{Abramowicz:1988sp,Narayan:2003by, Yuan:2014gma}, where the synchrotron polarization is set by the local magnetic-field geometry and flow structure \cite{M87_8,SgrA_8,EventHorizonTelescope:2021btj, Emami:2022kci}, providing a potential probe of spacetime geometry \cite{Ricarte:2022wpd, Hou:2024qqo}.
Simulations indicate that post-merger accretion emission can be substantial \cite{Ennoggi:2025ijc} and, for LISA-band sources, may last for hours. Such systems are thus promising targets for detecting ringdown imprints on polarization with current and upcoming facilities \cite{eXTP:2016rzs, weisskopf2022imaging, Ayzenberg:2023hfw, Johnson:2024ttr}.

More broadly, our result supports that electromagnetic polarimetry could provide an independent complement to gravitational-wave measurements of black-hole perturbations. Comparing time-resolved PA swings with variations in polarized flux and, where available, total intensity may help determine whether a candidate signal is consistent with a spacetime-driven polarization response, rather than with ordinary source variability.
 

\medskip

\begin{acknowledgments}
\noindent\emph{\textbf{Acknowledgments}}--- We thank Y. F. Chen, X. Chen, and Z. Li for helpful discussions. The work is partly supported by NSFC Grant No.12275004, 12205013, 12575048, 12547123. 
M. Guo is also supported by the BNU Tang Scholar. Zhen Zhong is supported by the MUR FIS2 Advanced Grant ET-NOW (CUP:~B53C25001080001) and by the INFN TEONGRAV initiative.
\end{acknowledgments}

\bibliographystyle{utphys}
\bibliography{refs}

\medskip

\appendix

\pagebreak
\setcounter{secnumdepth}{3} 
\setcounter{equation}{0}
\setcounter{figure}{0}
\setcounter{table}{0}
\renewcommand\thesection{\Alph{section}}
\renewcommand\thesubsection{\arabic{subsection}}
\renewcommand\theequation{\Alph{section}.\arabic{equation}} 
\renewcommand{\thefigure}{S\arabic{figure}}
\renewcommand{\thetable}{S\arabic{table}}
\renewcommand\theHsection{S\Alph{section}}
\renewcommand\theHsubsection{S\arabic{subsection}}
\renewcommand\theHequation{S\Alph{section}.\arabic{equation}}
\renewcommand\theHfigure{S\arabic{figure}}
\renewcommand\theHtable{S\arabic{table}}

\widetext
\begin{center}
\textbf{\large Supplementary Materials}
\end{center}

\section{ Derivation of the perturbative equations}\label{App:A}

To examine the impact of gravitational perturbations on light propagation and polarization transport, we work in a generally perturbed spacetime. Within linear perturbation theory, the full metric $\bar g_{\mu\nu}$ is decomposed into a background metric $g_{\mu\nu}$ and a first-order perturbation,
\begin{align}
\bar{g}_{\mu\nu}=g_{\mu\nu}+\epsilon\, h_{\mu\nu}\,,
\end{align}
where $\epsilon$ is a bookkeeping parameter that tracks the perturbative order, and $h_{\mu\nu}$ satisfies the linearized Einstein equations.
The associated Christoffel symbols and Riemann tensor are defined as
\begin{align}\label{eqapp:CR1}
    \tensor{\bar{\Gamma}}{^\mu_\nu_\rho}=\dfrac{1}{2}\bar{g}^{\mu\sigma}
    \pa{
       \partial_{\nu}\bar{g}_{\sigma\rho}+\partial_{\rho}\bar{g}_{\nu\sigma}-\partial_{\sigma}\bar{g}_{\nu\rho}
    }\,,
    \qquad
\tensor{\bar{R}}{^\mu_\alpha_\beta_\nu}=2
    \partial_{[\alpha} \tensor{\bar{\Gamma}}{^\mu_{\nu]}_\beta}+2\tensor{\bar{\Gamma}}{^\mu_\rho_{[\alpha}}\tensor{\bar{\Gamma}}{^\rho_{\nu]}_\beta}\,.
\end{align}
Expanding these quantities to first order in $\epsilon$, we obtain
\begin{align}
    \tensor{\bar{\Gamma}}{^\mu_\nu_\rho}
    =
    \tensor{{\Gamma}}{^\mu_\nu_\rho}+
    \epsilon\, {
    \tensor{H}{^\mu_\nu_\rho}}
    +\mathcal{O}\pa{\epsilon^2}\,,
    \qquad
    \tensor{\bar{R}}{^\mu_\alpha_\beta_\nu}=
    \tensor{R}{^\mu_\alpha_\beta_\nu}+\epsilon
    \,^{(1)}\tensor{{R}}{^\mu_\alpha_\beta_\nu}+
    \mathcal{O}\pa{\epsilon^2}\,,
\end{align}
where $\tensor{H}{^\mu_\nu_\rho}$ and $\,^{(1)}\tensor{{R}}{^\mu_\alpha_\beta_\nu}$ denote the linearized Christoffel symbols and Riemann tensor induced by $h_{\m\n}$, respectively. Although the Christoffel symbols themselves are not tensors, their linear perturbation $\tensor{H}{^\mu_\nu_\rho}$ transforms tensorially, as it can be written in the covariant form
\begin{align}
    \tensor{H}{^\mu_\nu_\rho}=\dfrac{1}{2}g^{\mu\sigma}
    \pa{
        \nabla_{\nu}h_{\sigma\rho}+\nabla_{\rho}h_{\nu\sigma}-\nabla_{\sigma}h_{\nu\rho}
    }\,,
\end{align}
where $\nabla_\m$ denotes the covariant derivative associated with the background metric.
In terms of $\tensor{H}{^\mu_\nu_\rho}$, the linearized Riemann tensor takes the compact form, $\,^{(1)}\tensor{{R}}{^\mu_\alpha_\beta_\nu}=2\nabla_{[\alpha} \tensor{H}{^\mu_{\nu]}_\beta}$.

\subsection{The geodesic deviation}
\begin{figure}[!htp]
    \centering
    \includegraphics[width=0.5\textwidth]{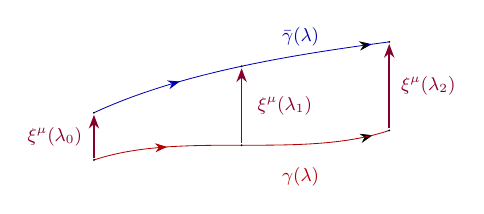}
      \caption{Gravitational perturbations modify the trajectories of light; the unperturbed and perturbed paths are denoted by $\gamma\pa{\lambda}$ and $\bar\gamma\pa{\lambda}$, respectively.}
    \label{fig:syt1}
\end{figure}

Within the geometric optics approximation, light rays follow null geodesics governed by
\begin{align}
    \dfrac{\dif}{\dif\lambda}
    \bar{k}^\mu=-\tensor{\bar{\Gamma}}{^\mu_\nu_\rho}
    \bar{k}^\nu \bar{k}^\rho\,,
    \label{eqn:geodesic_eqn}
\end{align}
where $\bar{k}^\mu=\dif \bar{x}^\mu/\dif \lambda$ is the photon four-momentum and $\lambda$ is an affine parameter. For weak perturbations induced, for instance, by typical gravitational waves, we expand the null geodesic (labeled by $\bar\gamma(\lambda)$) in powers of $\epsilon$ as
\begin{align}
    \bar{x}^\mu\pa{\lambda}=
    x^\mu\pa{\lambda}+\epsilon \xi^\mu \pa{\lambda}+\mathcal{O}\pa{\epsilon^2}\,,
    \label{eqn:zks1}
\end{align}
where $\xi^\mu(\lambda)$ is the deviation vector describing the displacement of the perturbed trajectory relative to the background geodesic, as illustrated in Fig.~\ref{fig:syt1}. Using Eqs.~\eqref{eqapp:CR1} and \eqref{eqn:zks1}, the Christoffel symbols evaluated along the perturbed trajectory can be expanded as
\begin{align}
    \tensor{\bar{\Gamma}}{^\mu_\nu_\rho}
    \pa{\bar{x}^\mu}=
    \tensor{{\Gamma}}{^\mu_\nu_\rho}\pa{x^\mu}+
    \epsilon \left[\xi^\alpha\partial_\alpha 
    \tensor{{\Gamma}}{^\mu_\nu_\rho}\pa{x^\mu}
    +\tensor{H}{^\mu_\nu_\rho}\pa{x^\mu}\right]
    +\mathcal{O}\pa{\epsilon^2}\,,
     \label{eqn:zks2}
\end{align}
where all quantities on the right-hand side are evaluated at $x^\m$.
Substituting Eqs.\eqref{eqn:zks1} and \eqref{eqn:zks2} into Eq.\eqref{eqn:geodesic_eqn} and collecting terms order by order in $\epsilon$, we obtain the equations governing light propagation. At leading order,
\begin{align}
    \dfrac{\dif}{\dif\lambda}
    {k}^\mu=-\tensor{{\Gamma}}{^\mu_\nu_\rho}
    {k}^\nu {k}^\rho\,.
    \label{eqn:bjskpxyd}
\end{align}
which determines the background null geodesic (denoted by $\gamma(\lambda)$). At sub-leading order, the deviation vector $\xi^\mu$ satisfies a second-order equation:
\begin{align}
    \dfrac{\dif^2}{\dif\lambda^2}\xi^\mu=
    & -\pa{\partial_\alpha 
    \tensor{{\Gamma}}{^\mu_\nu_\rho}
    \xi^\alpha+\tensor{H}{^\mu_\nu_\rho}}
    k^\nu k^\rho -
     2\tensor{{\Gamma}}{^\mu_\nu_\rho}  
     \dfrac{\dif\xi^\nu}{\dif\lambda}
     k^\rho \,,
     \label{eqn:yjfccdd}
\end{align}
which depends on the background connection, its derivatives, and the perturbative contribution induced by the gravitational wave. Combining it with Eq.~\eqref{eqn:bjskpxyd}, one can recast it into the compact covariant form
\begin{align}
   k^\alpha\nabla_\alpha(k^\beta\nabla_\beta \xi^\mu)+\tensor{R}{^\mu_\alpha_\nu_\beta}
       k^\alpha k^\beta \xi^\nu=-
        \tensor{H}{^\mu_\nu_\rho}k^\nu k^\rho\,,
\label{eqn:cdplfc}
\end{align}
where $k^\mu\nabla_\mu$ denotes the covariant derivative along the background geodesic $\gamma(\lambda)$. This equation incorporates both the tidal effects of the background spacetime and the direct forcing induced by gravitational perturbations in a fully covariant manner, generalizing the corresponding flat-spacetime results \cite{Damour:1998jm,Kopeikin:2001dz,Book:2010pf}.

\subsection{Separation equations for tetrad transport}

Vectors parallel transported along null geodesics can be treated transparently within the perturbative framework. To this end, we consider a general orthonormal tetrad $\{\bar e^\mu_{(a)}, a = 0,1,2,3\}$ defined along the perturbed null geodesic $\bar\gamma(\lambda)$, satisfying the parallel transport equation:
\begin{align}
\bar k^\alpha\bar\nabla_\alpha \bar e^\mu_{(a)}=0\,.
\label{eqn:pxydfcbj}
\end{align}
To isolate the effect of gravitational perturbations, we expand the tetrad as  
\begin{align}
    \bar e^\mu_{(a)}= e^\mu_{(a)}+\epsilon\, {e^\mu_1}_{(a)}
    +\mathcal{O}\pa{\epsilon^2}\,,
\end{align}
where $e^\mu_{(a)}$ is the background tetrad defined along $\gamma(\lambda)$, and ${e^\mu_1}_{(a)}$ denotes the first-order correction. 
Substituting this expansion into Eq.~\eqref{eqn:pxydfcbj} and matching orders in $\epsilon$, the leading-order equation yields the standard parallel transport in the background spacetime, $k^\alpha\nabla_\alpha e^\mu_{(a)}=0$. At first order, one obtains the evolution equation:
\begin{align}
   k^\alpha\nabla_\alpha {e^\mu_1}_{(a)}=
    & -\pa{\partial_\alpha 
    \tensor{{\Gamma}}{^\mu_\nu_\rho}
    \xi^\alpha+\tensor{H}{^\mu_\nu_\rho}}
    k^\nu e^\rho_{(a)}-
     \tensor{{\Gamma}}{^\mu_\nu_\rho}  
     \dfrac{\dif\xi^\nu}{\dif\lambda}
     e^\rho_{(a)} \,.
     \label{eqn:yjfccdd78jj}
\end{align}
Thus ${e^\mu_1}_{(a)}$ satisfies a first-order differential equation along the ray, sourced by the background connection, its derivatives, and the gravitational perturbation. 
However, $\bar e^\mu_{(a)}$ and $e^\mu_{(a)}$ are defined on different geodesics, $\bar\gamma(\lambda)$ and $\gamma(\lambda)$, which in turn lie in different spacetimes. As a result, ${e^\mu_1}_{(a)}$ and Eq.~\eqref{eqn:yjfccdd78jj} are not, by themselves, covariant and therefore do not directly correspond to observables. To restore covariance, we introduce a pullback that maps $\bar{e}^\mu_{(a)}$ from $\bar\gamma(\lambda)$ onto $\gamma(\lambda)$. Specifically, we have the new tetrad, $\widetilde{e}^{\mu}_{(a)}$, which is given by
\begin{align}
    \widetilde{e}^{\mu}_{(a)}=\bar{e}^\mu_{(a)}+\epsilon\pa{\tensor{\Gamma}{^\mu_\nu_\rho}\xi^\rho+\dfrac{1}{2}\tensor{h}{^\mu_\nu} }
    \bar{e}^\nu_{(a)}+\mathcal{O}\pa{\epsilon^2} \,,
    \label{eqn:tetrad_transformation}
\end{align}
where the term $\tensor{\Gamma}{^\mu_\nu_\rho}\bar{e}_{(a)}^\nu\xi^\rho$ accounts for parallel transport along the deviation vector $\xi^\mu$ from $\bar\gamma(\lambda)$ to $\gamma(\lambda)$, while $\tensor{h}{^\mu_\nu}\bar{e}_{(a)}^\nu$ arises from the identification between the perturbed and background metrics.  By construction, $\widetilde{e}^{\mu}_{(a)}$ satisfies the orthonormality condition
\begin{align}
   g_{\mu\nu}\pa{x} \widetilde{e}^{\mu}_{(a)}\widetilde{e}^{\nu}_{(a)}=\eta_{(a)(b)}+\mathcal{O}\pa{\epsilon^2}\,.
\end{align}
Using this new tetrad, we can safely analyze the physical imprint of perturbations along null rays. For clarity, we expand the tetrad as
\begin{align}
    \widetilde{e}^{\mu}_{(a)}={e}^{\mu}_{(a)}+\epsilon\, \delta e^\mu_{(a)}+\mathcal{O}\pa{\epsilon^2}
    \,.
\end{align}
Applying Eq.\eqref{eqn:tetrad_transformation}, we obtain $\delta e^\mu_{(a)}={e^\mu_1}_{(a)}+\left(\tensor{\Gamma}{^\mu^\nu_\rho}\xi^\rho+{1}/{2}\,\tensor{h}{^\mu_\nu}\right) {e}^\nu_{(a)}$. Substituting this relation into Eq.\eqref{eqn:yjfccdd78jj} yields the evolution equation
\begin{align}
   k^\alpha\nabla_\alpha\delta e^\mu_{(a)}
   =  e^\nu_{(a)}\tensor{R}{^\mu_\nu_\alpha_\beta}
      k^\alpha \xi^\beta+e^{\nu}_{(a)}  k^\rho \nabla_{[\mu} h_{\nu]\rho}  \,.
      \label{eqn:bzckfc1}
\end{align}
Since $e^\mu_{(a)}$ and $\widetilde{e}^\mu_{(a)}$ are now distinct orthonormal tetrads defined on the same spacetime, the linear perturbation $\delta e^\mu_{(a)}$ characterizes their relative rotation and thus corresponds to observables. To quantify this rotation, we project $\delta e^\mu_{(a)}$ onto the background tetrad and define the rotation angle in the $(a,b)$ plane as
\bea
\Theta_{(a)(b)} = g_{\mu\nu} \delta e^\mu_{(a)}e^\nu_{(b)} \,.
\eea
Using the completeness of the orthonormal tetrad, we introduce the antisymmetric tensor
\begin{align}
    \Theta_{\mu\nu}=\Theta_{(a)(b)} e_\mu^{(a)} e_{\nu}^{(b)}=\eta_{(a)(b)}\delta e_\mu^{(a)} e_{\nu}^{(b)}\,,
    \label{eqn:fczcdddsfd}
\end{align}
which fully encodes the perturbation-induced variations of the tetrad. Contracting Eq.~\eqref{eqn:bzckfc1} with $e^{(a)}_\nu$, we obtain the evolution equation for $\Theta_{\mu\nu}$:
\begin{align}
    k^\alpha\nabla_\alpha \Theta_{\mu\nu} = \tensor{R}{_\mu_\nu_\alpha_\beta}
        k^\alpha \xi^\beta +  k^\rho \nabla_{[\mu} h_{\nu]\rho}\,.
        \label{eqn:mian_function}
\end{align}
The first term on the right-hand side arises from geodesic deviation and encodes the background-curvature effect, while the second term captures the gravitational perturbations. Eq.~\eqref{eqn:mian_function} constitutes the central theoretical formula of this work, providing a transparent and covariant characterization of GW-induced polarization swing.

\section{Photon-polarization rotation}\label{App:B}

In this section we analyze the evolution of polarization along perturbed light rays, beginning with a general formulation in curved spacetimes. The photon polarization state is fully described by the polarization tensor $S^{\mu\nu}$, proportional to the phase-space density matrix \cite{Broderick:2003fc}. It is Hermitian, $S^{\mu\nu}=(S^{\nu\mu})^*$. To remove gauge freedom and isolate the physical Stokes parameters, we adopt the radiation gauge and expand $S^{\mu\nu}$ in an orthonormal tensor basis,
\begin{align}
S^{\mu\nu}=
\dfrac{1}{2}&\left(
\mathcal{I}\,e_{B}^{\mu\nu}+\mathcal{Q}\,e_{+}^{\mu\nu}+\mathcal{U}\,e_{\times}^{\mu\nu}+i\mathcal{V}\,e_{A}^{\mu\nu}
\right)\,,
\label{eq:S_expansion}
\end{align}
where $\mathcal{S} = \{\mathcal{I}, \mathcal{Q}, \mathcal{U}, \mathcal{V}\}$ are invariant Stokes parameters. The frequency-resolved intensities are $I_\n = \n^3\mathcal{I}$, $Q_\n = \n^3\mathcal{Q}$, $U_\n = \n^3\mathcal{U}$,  $V_\n = \n^3\mathcal{V}$, where $\n$ s the photon frequency. The basis tensors in Eq.~\eqref{eq:S_expansion} can be constructed by
\begin{align}
    {e}_{A}^{\mu\nu}=\xtd{\mu}\ytd{\nu}-\ytd{\mu}\xtd{\nu}\,,\qquad
    {e}_{B}^{\mu\nu}=\xtd{\mu}\xtd{\nu}+\ytd{\mu}\ytd{\nu}\,, \nonumber\\
    {e}_{+}^{\mu\nu}=\xtd{\mu}\xtd{\nu}-\ytd{\mu}\ytd{\nu}\,,\qquad
    {e}_{\times}^{\mu\nu}=\xtd{\mu}\ytd{\nu}+\ytd{\mu}\xtd{\nu}\,, \label{eqn:zljdzbdy}
\end{align}
where $e^\mu_{(X)}$, $e^\mu_{(Y)}$ are orthonormal vectors orthogonal to the wavevector $k^\mu$, spanning the transverse subspace. In vacuum, the polarization tensor is parallel transported, $ k^\alpha \nabla_\alpha S^{\mu\nu}=0$, which is equivalent to parallel transport of the tetrad and conservation of the Stokes parameters:
\begin{align}
    {k}^\nu \nabla_\nu \,e^\mu_{(a)}=0\,,
    \qquad 
    \dfrac{\dif}{\dif \lambda} \mathcal{S}=0\,.
    \label{eqn:dfddddddd}
\end{align}

To assess the impact of gravitational perturbations on photon polarization—from emission to propagation along the geodesic—we align the tetrads at the source by imposing $\widetilde{e}^\mu_{(a)}(\lambda_s)=e^\mu_{(a)}(\lambda_s)$. We denote the Stokes parameters before and after the perturbation by $\mathcal{S}$ and $\bar{\mathcal{S}}$, respectively. The polarization tensor transported in the perturbed spacetime is $\bar{S}^{\mu\nu}$. Its pullback to the background spacetime, given by Eq.\eqref{eqn:tetrad_transformation}, reads
\begin{align}
\widetilde S^{\mu\nu}=
\dfrac{1}{2}&\left(
\bar{\mathcal{I}}\,\widetilde e_{B}^{\mu\nu}+\bar{\mathcal{Q}}\,\widetilde e_{+}^{\mu\nu}+\bar{\mathcal{U}}\,\widetilde e_{\times}^{\mu\nu}+i\bar{\mathcal{V}}\,\widetilde e_{A}^{\mu\nu}
\right)\,.
\label{eq:S_expansion_pre}
\end{align}
Along the trajectory, the perturbation-induced rotation from $\widetilde{e}^\mu_{(a)}(\lambda)$ to $e^\mu_{(a)}(\lambda)$ is fully encoded in $\Theta_{\mu\nu}$ (Eq.~\eqref{eqn:fczcdddsfd}), such that$\widetilde{e}^\mu_{(a)}(\lambda)=e^\mu_{(a)}(\lambda)+\tensor{\Theta}{_{(a)}^{(b)}}e^\mu_{(b)}(\lambda)$. Expanding $\widetilde{S}^{\mu\nu}$ on the background basis then gives
\begin{align}
    \widetilde{S}^{\mu\nu}= &\dfrac{1}{2}\Big[
       \bar{\mathcal{I}}\,{e}_{B}^{\mu\nu} 
+\pa{\bar{\mathcal{Q}}-2\Theta_{(x)(y)} \bar{\mathcal{U}}}\,{e}_{+}^{\mu\nu}+
        \pa{\bar{\mathcal{U}}+2\Theta_{(x)(y)} \bar{\mathcal{Q}}}\,{e}_{\times}^{\mu\nu} 
+i\bar{\mathcal{V}}\,{e}_{A}^{\mu\nu}
    \Big]\,+\,\text{gauge terms}\,.
\end{align}
The gauge terms arise from rotations outside the $(x,y)$ plane and vanish upon projection onto the subspace transverse to $k^\mu$. Consequently, $\Theta_{(x)(y)}$ is the only component relevant for the physical polarization.

Changes in polarization intensities are determined solely by perturbations at the source, whereas the direction of linear polarization (the electric vector position angle, EVPA) depends on both source effects and propagation. The observed rotation,
\begin{align}
    \Delta\chi = \dfrac{1}{2}\atan2\pa{\bar{\mathcal{Q}}-2\Theta_{(x)(y)} \bar{\mathcal{U}} \,,
    \bar{\mathcal{U}}+2\Theta_{(x)(y)} \bar{\mathcal{Q}} }- \dfrac{1}{2}\atan2\pa{\mathcal{Q}\,,\mathcal{U}}\,,
\end{align}
measures the EVPA difference between the perturbed and background spacetimes. Here $\atan2(y,x)$ returns the angle of $(x,y)$ with the correct quadrant, reducing to $\arctan(y/x)$ when $x > 0$ but adjusting by $\pm \pi$ when $x < 0$ to resolve ambiguity. It is convenient to decompose $\Delta\chi=\Delta\chi_s+\vartheta$, with
\begin{align}
    \Delta\chi_{{s}}=\dfrac{1}{2}\pa{
    \atan2\pa{\bar{\mathcal{Q}},\bar{\mathcal{U}} }- \atan2\pa{\mathcal{Q},\mathcal{U}}}
    \,,\qquad
    \vartheta=\dfrac{1}{2}\tan^{-1}\pa{2\Theta_{(x)(y)}}\approx \Theta_{(x)(y)}\,.
    \label{eqn:PPRdef}
\end{align}
Here, $\Delta\chi_{s}$ captures the intrinsic rotation imprinted in the source region, while $\vartheta$ is the perturbed polarization rotation (PPR) accumulated during propagation. The latter depends only on the spacetime geometry along the ray and is independent of the initial polarization state. Combining this definition with Eq.~\eqref{eqn:mian_function}, the PPR evolves as
 \begin{align}
    \frac{\dif}{\dif\lambda} \vartheta =e^\mu_{(Y)}e^\nu_{(X)} \pa{\tensor{R}{_\mu_\nu_\alpha_\beta}
        k^\alpha \xi^\beta +  k^\rho \nabla_{[\mu} h_{\nu]\rho}}\,.
        \label{eqn:mian_function_of_PPR}
\end{align}
In the main text we retain only the PPR contribution to the observed EVPA, neglecting source-induced variations. This is justified because the PPR accumulates along the geodesic through the QNM region, whereas source effects arise from highly localized perturbations and are subdominant. Nevertheless, it is of interest to investigate in future work how the source-induced term depends explicitly on the perturbations and emission geometry.

\section{Time dependence of the PPR}
\label{app:timePPR}

Guided by the structure of Eq.~\eqref{eqn:mian_function_of_PPR}, we define the rotational angular velocity $\Omega$ as a covariant generalization of the angular velocity introduced in \cite{Kopeikin:2001dz}:
\begin{align}
      \Omega= e^\mu_{(Y)}e^\nu_{(X)} \, k^\rho \nabla_{[\mu} h_{\nu]\rho} 
      +e^\mu_{(Y)}e^\nu_{(X)} \tensor{R}{_\mu_\nu_\alpha_\beta}
        k^\alpha \xi^\beta 
      \,,
      \label{eq:omega_general}
\end{align}
This expression separates naturally into two contributions. The first term, $\Omega_h = e^\mu_{(Y)}e^\nu_{(X)} k^\rho \nabla_{[\mu} h_{\nu]\rho}$, arises directly from the metric perturbation. The second, $\Omega_\xi = e^\mu_{(Y)}e^\nu_{(X)} R_{\mu\nu\alpha\beta} k^\alpha \xi^\beta$, encodes rotation induced by geodesic deviation in the background curvature.
Accordingly, the total PPR accumulated along the ray from the source $x^\mu_s$ to the observer $x^\mu_o$ decomposes as $\vartheta = \vartheta_h + \vartheta_\xi$, with
\begin{align}
    \vartheta_h\pa{x_s^\mu,x_o^\mu}=\int_{\gamma\pa{\lambda}}   \dif\lambda\,\Omega_h \pa{x^\mu\pa{\lambda}}\,,
    \qquad
     \vartheta_\xi \pa{x_s^\mu,x_o^\mu}=\int_{\gamma\pa{\lambda}}   \dif\lambda\,\Omega_\xi \pa{x^\mu\pa{\lambda}}\,.
     \label{eq:werewrewvvv}
\end{align}
In our setup, the geodesic deviation vector $\xi^\mu$ is itself sourced by the perturbations, rendering $ \vartheta_\xi$ higher order. Moreover, radial curvature components typically decay faster than the metric perturbations, implying $ \vartheta_\xi \ll \vartheta_h$. This hierarchy is supported by the numerical results presented in the main text and is further corroborated in \ref{sub:nveps}.

\subsection{GW-induced rotation ($\vartheta_h$) }
In this section, we apply the general framework developed above to the ringdown phase of a Kerr black hole. We consider a single-frequency quasinormal mode (QNM), for which the metric perturbation takes the form
\begin{align}
    h_{\mu\nu}\pa{x^\mu}=\Re\left\{A_{\mu\nu}\pa{x^i}e^{-i\omega t}
    \right\} H\pa{t-r_*}\,,
    \label{eqn:dmsylb}
\end{align}
where $\omega = \omega_R - i \omega_I$ denotes the complex frequency, $A_{\mu\nu}(x^i)$ is the spatially dependent complex amplitude, and $H(t-r_*)$  is a Heaviside function that confines the perturbation to the interior of the light cone $t = r_*$, with $r_*$ the Kerr tortoise coordinate. Owing to the linearity of Eq.\eqref{eq:omega_general}, a time-harmonic metric perturbation induces a time-harmonic angular velocity $\Omega_h$:
\begin{align}
    \Omega_h\pa{x^\mu}=\Re\left\{\tilde{\Omega}\pa{x^i} e^{-i\omega t}\right\}H\pa{t-r_*}\,,
    \label{eq:PPR_general}
\end{align}
where $\tilde{\Omega}(x^i) = e^\mu_{(Y)} e^\nu_{(X)} k^\rho \left( \nabla_{[\mu} A_{\nu]\rho} - i \omega \delta_{[\mu}^0 A_{\nu]\rho} \right)$ is a complex function depending only on spatial coordinates. Substituting this decomposition into Eq.\eqref{eq:werewrewvvv}, we obtain
\begin{align}
    \vartheta_h=\Re\left\{\int_{\gamma\pa{\lambda}} \dif\lambda\,\tilde{\Omega} \pa{x^i\pa{\lambda}}e^{-i\omega t}H\pa{t({\lambda})-r_*({\lambda})}\right\}\,,
    \label{eqn:hfhjdsghjfsddfsa}
\end{align}
where the domain of integration is restricted by the step function. This expression shows that the time evolution of the observed PPR, as a function of the observation time $t_o$, can be obtained by tracing null geodesics that reach the observer at different arrival times.

To compute the PPR accumulated along a light ray arriving at the observer at coordinate time $t_o$, we must relate the coordinate time $t\pa{\lambda}$ at any point along the geodesic $\gamma(\lambda)$ to $t_o$.  In a stationary background, the propagation delay $\Delta t(\lambda)$ between a point $x^i(\lambda)$ and the observer $x^i_o$ iis independent of the arrival time. Hence, 
\begin{align}
t(\lambda) = t_o - \int \frac{\dif t}{\dif\lambda'} \dif \lambda' \equiv t_o-(r_*)_o - \Delta t(\lambda)\,.
\end{align}
For later convenience, we set $\Delta t(\lambda_o)=-(r_*)_o$ and introduce the shifted arrival time $\tilde{t}_o=t_o-(r_*)_o$. Substituting this relation into Eq.~\eqref{eqn:hfhjdsghjfsddfsa}, and noting that $e^{-i\omega \tilde{t}_o}$ is independent of $\lambda$, we factor it out of the integral to obtain
\begin{align}
\vartheta_h (\tilde{t}_o) = \Re \left\{ e^{-i\omega \tilde{t}_o} \int_{\gamma} \dif\lambda\,\tilde{\Omega}(x^i(\lambda)) e^{i\omega \Delta t(\lambda)} H(\tilde{t}_o - \tau(\lambda))   \right\}, \label{eq:PPR_final_integral}
\end{align}
where $\tau(\lambda) \equiv \Delta t(\lambda) + r_*(\lambda)$ defines an effective retarded-time threshold for the wavefront, satisfying $\tau(\lambda_o)=0$. It is convenient to introduce a time-dependent complex modulation factor, 
\begin{align}
\mathcal{A}(\tilde{t}_o) = \int_{\gamma}\dif \lambda \,\tilde{\Omega}(x^i(\lambda)) e^{i\omega \Delta t(\lambda)} H(\tilde{t}_o - \tau(\lambda))  \,,
\label{eqn:AAAdef}
\end{align}
in terms of which the observed PPR assumes the form of a damped harmonic oscillation,
\begin{align}
\vartheta_h(\tilde{t}_o) = |\mathcal{A}(\tilde{t}_o)|  \cos(\omega_R \tilde{t}_o-\Phi(\tilde{t}_o))e^{-\omega_I \tilde{t}_o}\,, \label{eq:theta_structure}
\end{align}
where $|\mathcal{A}(\tilde{t}_o)|$ and $\Phi(\tilde{t}_o)$ denote the magnitude and phase of $\mathcal{A}(\tilde{t}_o)$, respectively.

Exploiting the properties of null geodesics in Kerr spacetime, we consider two representative scenarios in what follows. The first involves photons emitted from the near-horizon region that propagate directly to a distant observer. The second describes scattering trajectories, in which photons originating at spatial infinity are deflected by the strong gravitational field before reaching a far-field observer.

{\textbf{Case \uppercase\expandafter{\romannumeral 1}:}} \par
For photons emitted from a source at \(r_s\) and propagating directly to an observer at \(r_o \to \infty\), the function \(\tau(\lambda)\) is monotonically increasing along the geodesic. Let \(\tau_s = \tau(\lambda_s)\) denote the effective retarded arrival time associated with the source. The evolution of \(\vartheta_h(\tilde{t}_o)\) can then be naturally divided into three distinct phases:
\begin{itemize}
    \item \textbf{Phase 1: Quiescent (\(\tilde{t}_o \le  0\))} \\
    In this regime, the GW front has not yet reached any segment of the geodesic that can causally affect the observer at time \(\tilde{t}_o\). Accordingly, the integral \(\mathcal{A}(\tilde{t}_o)\) vanishes, and one has \(\vartheta_h(\tilde{t}_o) = 0\).
    
    \item \textbf{Phase 2: Transient Growth (\(0 < \tilde{t}_o \le \tau_s\))} \\
    The Heaviside function activates a portion of the geodesic in the vicinity of the observer. The relevant integration domain becomes \([\lambda^*(\tilde{t}_o), \lambda_o]\), where \(\lambda^*(\tilde{t}_o)\) is defined implicitly by the causality condition \(\tau(\lambda^*(\tilde{t}_o)) = \tilde{t}_o\). The amplitude is therefore given by
    \begin{equation}
        \mathcal{A}(\tilde{t}_o) = \int_{\lambda^*(\tilde{t}_o)}^{\lambda_o} \tilde{\Omega}(x^i(\lambda)) e^{-i\omega \Delta t(\lambda)}\, \dif \lambda.
    \end{equation}
    As \(\tilde{t}_o\) increases, the point \(\lambda^*(\tilde{t}_o)\) moves from the observer toward the source, progressively enlarging the portion of the trajectory influenced by the perturbation. Consequently, the observed PPR amplitude increases monotonically, consistent with the behavior shown in the blue region of Fig.~\ref{fig:mdcx}.
    
    \item \textbf{Phase 3: Resonance Damping (\(\tilde{t}_o > \tau_s\))} \\
    Once the entire photon trajectory is immersed in the ringdown field, the integral \(\mathcal{A}(\tilde{t}_o)\) approaches a constant complex value \(\mathcal{A}\). The rotation angle then reduces to a purely damped harmonic form,
    \begin{align}
\vartheta_h(\tilde{t}_o) = |\mathcal{A}|  \cos(\omega_R\tilde{t}_o-\Phi)e^{-\omega_I \tilde{t}_o}\,,
\end{align}
in agreement with the PPR behavior shown in the red region of Fig.~\ref{fig:mdcx}.
\end{itemize}

{\textbf{Case  \uppercase\expandafter{\romannumeral 2}:}} \par

For deflected photon trajectories originating from a distant source (\(r_s \to \infty\)) and passing through a radial turning point \(r_t\) before reaching a distant observer, the geodesic exhibits a reversal in the radial direction. This structure allows the integral of the complex modulation factor to be decomposed into two contributions associated with the ingoing and outgoing segments of the trajectory, respectively, such that \(\mathcal{A}(\tilde{t}_o)=\mathcal{A}_{\mathrm{in}}(\tilde{t}_o)+\mathcal{A}_{\mathrm{out}}(\tilde{t}_o)\). Let \(\lambda_t\) denote the affine parameter at the turning point. The two contributions can then be written as
\begin{align}
\mathcal{A}_{\mathrm{in}}(\tilde{t}_o ) = \int^{\lambda_{t}}_{\lambda_s} \dif \lambda \,\tilde{\Omega}(x^i(\lambda)) e^{i\omega \Delta t(\lambda)} H(\tilde{t}_o - \tau(\lambda))  \,,
\quad
\mathcal{A}_{\mathrm{out}}(\tilde{t}_o ) =\int_{\lambda_{t}}^{\lambda_o}\dif \lambda \,\tilde{\Omega}(x^i(\lambda)) e^{i\omega \Delta t(\lambda)} H(\tilde{t}_o - \tau(\lambda))\,,
\end{align}
corresponding to the ingoing and outgoing branches of the photon path. We further define \(\tau_t = \tau(\lambda_t)\) as the effective retarded arrival time associated with the turning point.


\begin{itemize}
 \item \textbf{Phase 1: Quiescent \& Phase 2: Transient Growth ($\tilde{t}_o \le \tau_t$)} \\
 \par 
At early times, the interaction with the gravitational wave is entirely confined to the outgoing branch of the geodesic. In this regime, the ingoing contribution vanishes, i.e., $\mathcal{A}_{\mathrm{in}}(\tilde{t}_o) = 0$, and the total modulation is determined solely by the outgoing segment. The resulting evolution therefore reproduces the quiescent and transient growth phases identified in \textbf{Case \uppercase\expandafter{\romannumeral 1}}.

For $\tilde{t}_o > \tau_t$, the outgoing branch becomes fully immersed in the gravitational-wave field, and $\mathcal{A}_{\mathrm{out}}(\tilde{t}_o)$ asymptotes to a constant value $\mathcal{A}_{\mathrm{out}}$. Meanwhile, the ingoing branch progressively enters the perturbed region, leading to a monotonic growth of $\mathcal{A}_{\mathrm{in}}(\tilde{t}_o)$. Consequently, the signal is initially dominated by the outgoing contribution and gradually transitions to being governed by the ingoing branch at later times.
\end{itemize}

\begin{itemize}
        
    \item \textbf{Phase 3: Resonance Damping} \\
    In this regime, the signal is dominated by the outgoing-branch contribution, for which the quasinormal-mode response reaches its peak amplitude. The observer therefore measures prominent damped oscillations.

    \item \textbf{Phase 4: Late-time Power-law Tail ($\tilde{t}_o \to \infty$)} \\
    At sufficiently late times, the outgoing contribution becomes exponentially suppressed and hence subdominant, while the ingoing branch controls the asymptotic behavior. For rays originating at infinity, causality implies that, at large $\tilde{t}_o$, the photon enters the gravitational-wave-perturbed region at a radius $r \simeq \tilde{t}_o/2$. In the far-field limit, where $\dif \lambda \approx \dif t \approx -\dif r$, the integrand admits an expansion in inverse powers of $r$, with leading-order behavior
    \begin{align}
        \tilde{\Omega}(x^i) e^{i\omega \Delta t}\approx
        \frac{\mathcal{C}\pa{\theta,\phi}}{r^k} e^{i 2\omega r}\,,
    \end{align}
    where $\mathcal{C}(\theta,\phi)$ is a complex angular profile encoding the gravitational-wave amplitude, and $k$ is set by its multipolar structure. The ingoing contribution then evaluates to
    \begin{align}
        \mathcal{A}_{\mathrm{in}}(\tilde{t}_o ) \approx \int^{\tilde{t}_o/2}_{r_t} \dif r \, \frac{\mathcal{C}\pa{\theta,\phi}}{r^k} e^{i 2\omega r}
        \approx
        \left. \dfrac{\mathcal{C}\pa{\theta,\phi}}{i 2\omega}
        \frac{ e^{i 2\omega r}}{ r^k} \right|^{\tilde{t}_o/2}_{r_t} + \mathcal{O}(\dfrac{1}{\tilde{t}_o^{\, k+1}}) 
         \approx \dfrac{\mathcal{C}\pa{\theta,\phi}}{i 2^{1-k}\omega}
        \frac{ e^{i \omega \tilde{t}_o}}{ \tilde{t}_o^k}\,,
    \end{align}
    where the asymptotic expression follows from integration by parts in the limit $\tilde{t}_o \gg 1$. Since $|e^{i 2\omega r}| = e^{2\omega_I r}$ grows exponentially, the dominant contribution is the term proportional to $e^{i \omega \tilde{t}_o}$, yielding
    \begin{equation}
        \vartheta_h(\tilde{t}_o)   \approx  \Re\{\mathcal{A}_{\mathrm{in}}(\tilde{t}_o ) e^{-i\omega \tilde{t}_o } \}\sim \tilde{t}_o^{-k}.
        \label{eq:tail_derivation_dcvx}
    \end{equation}
\end{itemize}

The transition from exponential decay to a power-law tail has a clear geometric origin rooted in the structure of the geodesic integral. This late-time behavior provides a distinctive observable signature, effectively encoding the multipolar index $k$ of the black-hole perturbation in the far zone—information that is not directly accessible through standard quasinormal-mode spectroscopy.

\subsection{Curvature-deviation induced rotation ($\vartheta_\xi$)}
To evaluate the contribution of the term $\Omega_\xi$, one must determine the evolution of the deviation vector $\xi^\mu(\lambda)$, which satisfies the inhomogeneous geodesic deviation equation
\begin{align}
    \frac{D^2 \xi^\mu}{\dif \lambda^2} + \tensor{K}{^\mu_\nu}(\lambda)\,\xi^\nu = \mathcal{F}^\mu(\lambda)\,, \label{eq:Jacobi_inhomo}
\end{align}
where $D/\dif \lambda \equiv k^\alpha \nabla_\alpha$ denotes the covariant derivative along the background geodesic, $\tensor{K}{^\mu_\nu} = \tensor{R}{^\mu_\alpha_\nu_\beta} k^\alpha k^\beta$ is the tidal tensor, and the inhomogeneous driving term is $\mathcal{F}^\mu = - \tensor{H}{^\mu_\nu_\rho} k^\nu k^\rho$. The general solution to Eq.~\eqref{eq:Jacobi_inhomo} can be constructed in terms of two Jacobi propagators, $\tensor{S}{^\mu_\nu}(\lambda,\lambda')$ and $\tensor{C}{^\mu_\nu}(\lambda,\lambda')$, which serve as Green’s functions for the associated homogeneous system and satisfy
\begin{align}
    \frac{D^2}{\dif \lambda^2} \tensor{S}{^\mu_\nu}(\lambda, \lambda') + \tensor{K}{^\mu_\rho}(\lambda)\, S^\rho_{\ \nu}(\lambda, \lambda') = 0 \,,
    \qquad 
    \frac{D^2}{\dif \lambda^2} \tensor{C}{^\mu_\nu}(\lambda, \lambda') + \tensor{K}{^\mu_\rho}(\lambda)\, C^\rho_{\ \nu}(\lambda, \lambda') = 0 \,,
\end{align}
subject to the boundary conditions
\begin{align}
   \tensor{S}{^\mu_\nu}(\lambda', \lambda') = 0, 
    \quad
    \frac{D \tensor{S}{^\mu_\nu}}{\dif\lambda}\bigg|_{\lambda = \lambda'} = \tensor{\delta}{^\mu_\nu}\,,
    \qquad
     \tensor{C}{^\mu_\nu}(\lambda', \lambda') = \tensor{\delta}{^\mu_\nu},
    \quad
   \frac{D \tensor{C}{^\mu_\nu}}{\dif\lambda}\bigg|_{\lambda = \lambda'} = 0\,.
\end{align}
The deviation vector then admits the representation
\begin{equation}
\xi^\mu(\lambda) = \tensor{C}{^\mu_\alpha}(\lambda, \lambda_0)\, \xi^\alpha(\lambda_0)
+ \tensor{S}{^\mu_\alpha}(\lambda, \lambda_0)\, \frac{D\xi^\alpha}{\dif \lambda}\bigg|_{\lambda_0}
+ \int_{\lambda_0}^\lambda \tensor{S}{^\mu_\alpha}(\lambda, \lambda') \mathcal{F}^\alpha(\lambda') \,\dif \lambda'\,.
\end{equation}
Imposing the boundary conditions $\xi^\mu(\lambda_o) = 0$ and $D\xi^\mu/\dif \lambda|_{\lambda=\lambda_o} = 0$ at a distant observer, the solution is entirely sourced by the inhomogeneous term associated with the gravitational perturbation. Defining the background curvature operator $\mathcal{R}_\beta = e^\mu_{(Y)} e^\nu_{(X)} k^\alpha \tensor{R}{^\mu_\nu_\alpha_\beta}$ and substituting the above expression for $\xi^\mu$ into the definition of $\Omega_\xi$, one obtains
\begin{align}
    \Omega_\xi(\lambda) = \mathcal{R}_\alpha(\lambda) \int_{\lambda_o}^{\lambda} S^\alpha_{\ \beta}(\lambda, \lambda') \mathcal{F}^\beta(\lambda') \, \dif \lambda' \,.
\end{align}
In contrast to the local driving term $\Omega_h$, the quantity $\Omega_\xi$ captures nonlocal effects that depend on the entire past history of perturbations encountered along the photon trajectory. Its contribution to the total rotation angle therefore takes the form of a double integral,
\begin{align}
    \vartheta_\xi = \int_{\gamma} \dif \lambda \, \mathcal{R}_\alpha(\lambda) \int_{\lambda_o}^{\lambda} \tensor{S}{^\alpha_\beta}(\lambda, \lambda') \mathcal{F}^\beta(\lambda') \, \dif \lambda'\,.
    \label{eqn:dfhsjghudfshilufdsakgijuyfghius}
\end{align}
For a monochromatic gravitational wave described by Eq.~\eqref{eqn:dmsylb}, linearity implies that the driving term $\mathcal{F}^\mu$ can be written as
\begin{align}
    \mathcal{F}^\mu(x^\mu) = \Re\left\{\tilde{\mathcal{F}}^\mu(x^i)\, e^{-i\omega t}\right\} H(t - r_*)\,,
    \label{eq:PPR_general_F}
\end{align}
Substituting this expression into $\vartheta_\xi$ and proceeding analogously to the derivation of Eq.~\eqref{eq:PPR_final_integral}, one finds
\begin{align}
\vartheta_\xi (\tilde{t}_o) = \Re \left\{\mathcal{B}(\tilde{t}_o) e^{-i\omega \tilde{t}_o}\right\}
= |\mathcal{B}(\tilde{t}_o)| \cos\!\big(\omega_R \tilde{t}_o - \Psi(\tilde{t}_o)\big)\, e^{-\omega_I \tilde{t}_o}\,,
\label{eqn:fscdhjghjfdshjsfdah}
\end{align}
where the complex amplitude $\mathcal{B}(\tilde{t}_o) = |\mathcal{B}(\tilde{t}_o)| e^{i\Psi(\tilde{t}_o)}$ is defined by
\begin{align}
    \mathcal{B}(\tilde{t}_o) = \int_{\gamma} \dif \lambda \, \mathcal{R}_\alpha(\lambda)
    \int_{\lambda_o}^{\lambda} \dif \lambda' \, \tensor{S}{^\alpha_\beta}(\lambda, \lambda') \mathcal{F}^\beta(\lambda') \,
    e^{i\omega \Delta t(\lambda')} H(\tilde{t}_o - \tau(\lambda')) \,.
    \label{eqn:BBBBdef}
\end{align}
Since Eq.~\eqref{eqn:BBBBdef} is structurally analogous to Eq.~\eqref{eqn:AAAdef}, the rotation angle $\vartheta_\xi(\tilde{t}_o)$ exhibits qualitatively similar behavior to the cases discussed above. In particular, when the dominant photon trajectories are fully immersed in the gravitational-wave region—corresponding to the resonance damping phase—the Heaviside function evaluates to unity, removing the temporal cutoff. As a result, the complex amplitude $\mathcal{B}(\tilde{t}_o)$ approaches a constant $\mathcal{B}$, and $\vartheta_\xi(\tilde{t}_o)$ reduces to a damped harmonic oscillation.

\section{Numerical Validation: Exact vs. Perturbative Solutions}
\label{app:nveps}
This section validates the covariant perturbative framework through a direct comparison with fully numerical solutions. The latter are obtained by integrating the complete set of null geodesic and polarization parallel-transport equations in the dynamically perturbed spacetime, without invoking linearized expansions of either the tetrad or the geodesic deviation. The numerical evolution employs a 7th–8th order Runge–Kutta scheme, with the perturbed Kerr metric constructed via the Teukolsky–CCK formalism.

\begin{figure}[!htp]
    \includegraphics[width=\textwidth]{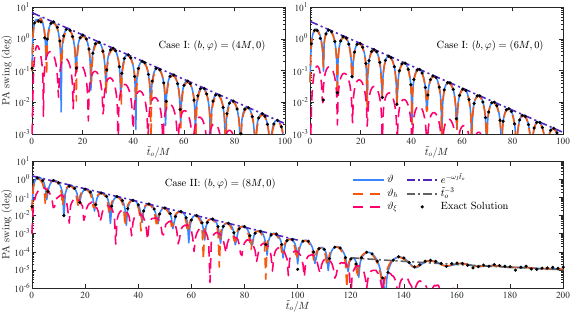}
\caption{Numerical validation of the covariant perturbative framework. The observer is located at spatial infinity with angular coordinates $\pa{\theta_o,\phi_o}=\pa{\pi,0}$. \textbf{Top}: Photons emitted from the equatorial plane \textbf{(Case I)} with impact parameters $b=4M$ (left) and $b=6M$ (right). \textbf{Bottom}: Photons originating from spatial infinity \textbf{(Case II)} with $b=8M$. In all panels, the absolute amplitude of the perturbed polarization rotation (PPR) is shown as a function of observer time $\tilde{t}_o$. The total perturbative prediction $\vartheta$ (solid blue) is in excellent agreement with the exact numerical solution (black dots) obtained from the full transport equations. The total PPR is further decomposed into the direct GW-induced rotation $\vartheta_h$ (dashed orange) and the curvature-deviation-induced rotation $\vartheta_\xi$ (dashed red). The clear dominance of $\vartheta_h$ over $\vartheta_\xi$ across all dynamical stages provides strong justification for the approximation $\vartheta \approx \vartheta_h$ adopted in the analytical treatment. Dash-dotted curves indicate the theoretical envelopes corresponding to exponential ringdown damping and the late-time power-law tail.
}
        \label{fig:compare1_of_two_model}
\end{figure}

Figure~\ref{fig:compare1_of_two_model} evaluates the accuracy of the perturbative framework for two representative emission configurations: an equatorial source (\textbf{Case \uppercase\expandafter{\romannumeral 1}}) and a source located at infinity (\textbf{Case \uppercase\expandafter{\romannumeral 2}}). The perturbative predictions exhibit excellent agreement with the exact results, accurately reproducing the oscillation frequency, damping rate, and phase of the polarization-angle evolution. Notably, this level of agreement persists even for trajectories passing in close proximity to the black hole.

We further decompose the total PPR $\vartheta$ into two components: the direct GW-induced contribution $\vartheta_h$ and the curvature-deviation-induced contribution $\vartheta_\xi$. As shown in Fig.~\ref{fig:compare1_of_two_model}, the local driving term $\vartheta_h$ dominates the total signal by at least an order of magnitude at all times. By contrast, the nonlocal term $\vartheta_\xi$, arising from curvature-coupled deviations, remains subdominant and decays more rapidly. These results provide strong numerical evidence that the observed PPR is primarily a direct imprint of the dynamical gravitational perturbation, rather than a secondary effect mediated by perturbations to the background lensing geometry.

\section{Parameter Dependence of the Polarization Response}
\label{sec:dfhhdj}
\begin{figure}
    [!htp]
    \includegraphics[width=0.49\textwidth]{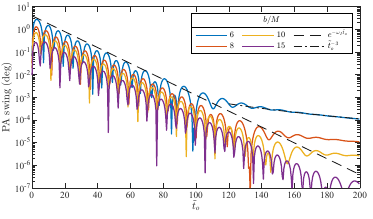}
    \includegraphics[width=0.49\textwidth]{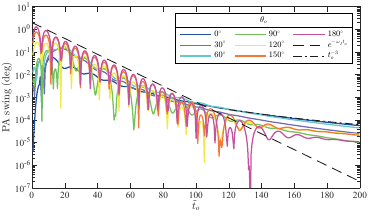}
     \caption{\textbf{Left}: Effect of ringdown on the polarization of light rays arriving from infinity for a range of impact parameters, with the observer inclination fixed at $\theta_o = \pi$. \textbf{Right}: Corresponding results for different inclination angles, with the impact parameters fixed at $\pa{b,\varphi}=\pa{8,0}$.}
        \label{fig:inf2ffrfrfrf}
\end{figure}

We present a systematic analysis of the polarization response for background sources located at infinity across a range of parameter configurations. Figure~\ref{fig:inf2ffrfrfrf} displays the temporal evolution of the observed PPR for different values of the impact parameter $b$ and observer inclination angle $\theta_o$. Across all cases, two characteristic features consistently emerge, in agreement with the analytical predictions for \textbf{Case II}. 

The first feature is a phase of rapidly damped oscillations during the ringdown stage ($0 \lesssim \tilde{t}_o \lesssim 100M$), where the polarization angle evolution closely tracks the fundamental quasinormal mode frequency and its associated exponential decay. The second feature is the onset of a distinct power-law tail at late times ($\tilde{t}_o \gtrsim 100M$), signaling the transition to the asymptotic regime of the perturbation.

\end{document}